\documentclass[useAMS,usenatbib]{mn2e}
\newcommand{\apj}{ApJ}
\newcommand{\apjl}{ApJL}
\newcommand{\mnras}{MNRAS}
\newcommand{\aap}{A\&A}
\newcommand{\ssr}{Space Science Reviews}
\usepackage{natbib}
\usepackage{amsmath}
\usepackage{amssymb}
\usepackage{graphicx}
\usepackage{color}
\usepackage{multirow}
\usepackage{color}
\usepackage{enumerate}
\usepackage[table]{xcolor}
\usepackage{float}
\usepackage{comment}
\usepackage{enumerate}
\usepackage{adjustbox}
\usepackage{subfigure}
\begin{document}
\title[Propeller effect on PSR B1259-63]{On the timing behavior of PSR B1259-63 under the propeller torque from a transient accretion disc}
\author[S. X. Yi et al.]{Shu-Xu Yi$^{\dagger}$ , K.S. Cheng$^{\dagger}$ \\
$^{\dagger}$ Department of Physics, the University of Hong Kong University, Pokfulam Road, Hong Kong\\
}
\maketitle
\begin{abstract}
The $\gamma$-ray pulsar binary system PSR B1259-63 flares in GeV after each periastron. The origin of those flares is still under debate. Recently, \cite{2017ApJ...844..114Y} proposed a mechanism that might explain the GeV flares. In that model, a transient accretion disc is expected to be formed from the matter which was gravity-captured by the neutron star from the main sequence companion's circumstellar disc. The transient accretion disc exerts a spin down torque on the neutron star (propeller effect), which might be traceable via pulsar timing observation of PSR B1259-63. In this paper, we phenomenologically consider the propeller effect with a parameter $\chi$, which describes the coupling between the disc matter and the neutron star. Comparing the expected timing residuals against the recent observation in \cite{2014MNRAS.437.3255S}, we conclude that the angular momentum transfer is very weak (with the coupling parameter $\chi\le10^{-4}$). 
\end{abstract}
\section{introduction}
PSR B1259-63/LS2883 is a close binary system, which is composed of a millisecond pulsar (PSR B1259-63) and a massive Be star (LS2883). The Be star rotates rapidly and develops a decretion disc (circumstellar disc, CD) in its equatorial plane. Since the orbit of the pulsar is highly eccentric, with the eccentricity $e=0.87$, and the nearest distance between the pulsar and the Be star is very small, $\sim0.9$\,A.U. \citep{2011ApJ...732L..11N}, the radio emission from the pulsar is eclipsed by the CD during $\sim T_0\pm15$\,days \citep{1996MNRAS.279.1026J}, where $T_0$ is the epoch of periastron. The interaction between the pulsar wind and the stellar wind (and/or the CD) gives rise to emission in X-ray and TeV bands \citep{1999ApJ...521..718H,1999APh....10...31K,2005A&A...442....1A,2006MNRAS.367.1201C,2014MNRAS.439..432C}. 

The binary is also found to be a GeV source with the {\it Fermi} satellite \citep{2011ApJ...736L..11A, 2011ApJ...736L..10T}. The GeV emission is quit puzzling: it remains quiescent through the orbit before a sudden flare at $\sim T_0+30$ days, where the pulsar is far away from the CD, and then gradually fades away in the following $\sim15$ days. The mechanism of the GeV emission is still under debate \citep{2012ApJ...752L..17K,2012ApJ...753..127K,2013A&A...557A.127D,2013ApJ...776...40M}. With multi-wavelength observations of the system around the 2014 periastron passage, \cite{2015MNRAS.454.1358C} found that the onset of the GeV flare to coincide with the rapid decay of the H$\alpha$ equivalent width, which favors the model proposed by \cite{2014MNRAS.439..432C}. Recently, we proposed a new model to account for the GeV flare of this system (\citealt{2017ApJ...844..114Y}, Y17). In that model, matter from the CD is transferred to the gravity capture radius of PSR B1259-63 and a transient accretion disc is developed around the neutron star. 

If the transient accretion disc does appear in each orbit, we shall expect the transfer of angular momenta between the accretion disc and the neutron star. In cases where the Keplerian velocity at the inner edge of the accretion disc is less than the co-rotating velocity with the neutron star, the accreted matter is stopped and ejected by the centrifugal force, which is known as the propeller effect \citep{1975A&A....39..185I}. As a result, the pulsar experiences a spin down torque. We will leave the introduction to the propeller torque in the next section. 

Timing observation of PSR B1259-63 is a natural way to study the potential time-varying braking torque from the transient disc. In the work of \cite{1995ApJ...445L.137M}, the authors found that the timing data of this pulsar can only be fitted after a jump of spin frequency included in the model of spin evolution (also known as the timing solution) at each periastron. The authors suggested that it was a clue of the propeller spin down. \cite{1998MNRAS.298...67W} found a timing solution other than that of \cite{1995ApJ...445L.137M}, in which the author included derivatives of the longitude of periastron passage ($\omega$) and the projected pulsar semi-major axis ($x$). Those derivatives originate from the quadrupole gravitational moment of the Be star \citep{1995ApJ...452..819L}. \cite{2004MNRAS.351..599W} applied three different timing solutions on the pulse time-of-arrival (TOAs) of PSR B1259-63: I, with $\dot{\omega}$ and $\dot{x}$ included; II, with jumps of the spin frequency and of its first derivative included at each periastron ($\Delta\nu$ and $\Delta\dot{\nu}$); III, with jumps of $x$ at each periastron included ($\delta x$). Solutions II and III result in significantly smaller root-mean-squares (rms) in the timing residuals than I, and III was preferred for its less degrees of freedom. Using more sophisticated orbit-spin coupling model, \cite{2014MNRAS.437.3255S} (S14) fitted the TOAs such that the residuals showed no orbital modulated timing residuals.

Since the timing solution of S14 fits the TOAs of PSR B1259-63 so well without requiring additional orbital modulated braking torques, their work can be used to set a limit on the transient disc scenario. In other words, if the propeller effect of the transient accretion disc is significant enough, any timing solution without considering it can not fit the TOAs so well. In this paper, we study the resulted timing residuals of the pulsar if the transient disc proposed in Y17 does appear around the pulsar after each periastron. Under the constrain from S14, the Y17 scenario is strongly limited.

\section{The braking torque from propeller effect}
The spin down rate of a pulsar which undergoes the propeller phase is composed with two parts:
\begin{equation}
\dot{\nu}=\dot{\nu}_{\rm{prop}}+\dot{\nu}_{\rm{dip}},
\end{equation}
where $\dot{\nu}_{\rm{prop}}$ is due to propeller effect, and $\dot{\nu}_{\rm{dip}}$ is due to the magnetic dipole radiation. 
The spin down torque exerted by the accretion disc can be calculated directly with:
\begin{equation}
N_{\rm{prop}}=-\int^\infty_{r_{\rm{M}}}r^2B_\phi B_zdr,
\label{eqn:realtorque}
\end{equation}
both analytically (see \citealt{1979ApJ...234..296G} and references hereafter) or numerically (see \citealt{2004ApJ...616L.151R} and references hereafter),where $B_\phi$ is the azimuthal magnetic field induced by the disc matter, $B_z$ is the vertical magnetic field component that penetrates the disc, $r_{\rm{M}}$ is the inner edge of the accretion disk. Many other papers estimate the spin down torque indirectly, by calculating the rate at which angular momentum is taken away by the ejected matter from the accretion disc (e.g., \citealt{1999ApJ...520..276M, 2014RAA....14...85L}):
\begin{equation}
2\pi I\dot{\nu}_{\rm{prop}}=2r^2_{\rm{M}}\dot{M}_{\rm{acc}}\left(\sqrt{\frac{GM_\star}{r^3_{\rm{M}}}}-2\pi\nu\right)\chi,\label{eqn:torque}
\end{equation}
where $I$, $\nu$, $M_\star$ and $\dot{M}_{\rm{acc}}$ are the rotational inertia, the spin frequency, the mass of the pulsar and the accretion rate of the disc at the inner most radius, respectively; 
$\chi\,(0<\chi<1)$ is the coefficient describing the degree of coupling between the accretion disc and the pulsar via the magnetic field lines. The factor 2 in the right hand side of the above equation comes because there are two nearly equal contributions of the torque: the angular momenta transferred at the inner edge of the disc and the angular momenta transferred from the accretion flow to the magnetic field beyond the inner edge \citep{1999ApJ...520..276M}. 

The link between the two approaches gives \citep{2006A&A...451..581D}: 
$2\chi=\sqrt{2}\gamma_{\rm{a}}\delta,$
where $\gamma_{\rm{a}}\sim B_\phi/B_z$ and $\delta\ll1$ is the ratio between the width of boundary layer of accretion and the inner radius of the accretion disc \citep{1979ApJ...232..259G}. $B_\phi$ is generated by rotation shear. In the case that the magnetosphere is nearly force free, the azimuthal pitch $\gamma_{\rm{a}}$ should be in the order of unity (see \citealt{1990A&A...227..473A,1992MNRAS.259P..23L}). As a result, $\chi\ll1$ is implied, which is in accordance with our findings below. 

In order to avoid dealing with the complexity and uncertainty of the magnetic field structure, we adopt equation (\ref{eqn:torque}) here. $r_{\rm{M}}$ is supposed to be equal to the Alfv\'en radius, where the ram pressure of the accretion flow is balanced by the magnetic field pressure. Then 
\begin{equation}
r_{\rm{M}}=5.1\times10^8\dot{M}_{\rm{acc,16}}^{-2/7}m_\star^{-1/7}\mu_{30}^{4/7}\,\text{cm}.
\label{eqn:alfven}
\end{equation}
$\dot{M}_{\rm{acc,16}}$ is the accretion rate of the accretion disc in units of $10^{16}$\,g/s, $\mu_{30}$ is the magnetic dipole in units of $10^{30}$\,G\,cm$^3$, $m_\star\equiv M_\star/M_\odot$.

If we assume the accretion disc to be the standard thin disc \citep{1973A&A....24..337S}, then:
\begin{equation}
\rho=3.1\times10^{-8}\alpha^{-7/10}\dot{M}^{11/20}_{\rm{acc},16}m_\star^{5/8}R_{10}^{-15/8}\,\rm{g/cm^3},\label{eqn:3}
\end{equation} 
where $\alpha$ is the viscosity index, $R_{10}$ is the radius in unit of $10^{10}$\,cm. 

The scenario where the transient disc is formed from the gravity-captured materials is similar to the case where a fossil disc is formed from fall-back materials around a newly born neutron star. Therefore we apply the equations from \cite{2000ApJ...534..373C} to describe the evolution of the accretion rate at the inner edge of the disc: 
\begin{equation}
\dot{M}_{\rm{acc}}=\dot{M}_{\rm{acc,0}}-\dot{M}_{\rm{eva}}\quad,\quad(r_{\rm{in}}>r_{\rm{M}})\nonumber
\end{equation}

\begin{equation}
\dot{M}_{\rm{acc}}=\dot{M}_{\rm{acc,0}}\big(\frac{t}{\tau}\big)^{-\beta}-\dot{M}_{\rm{eva}}\quad,\quad(r_{\rm{in}}=r_{\rm{M}})
\label{eqn:accrate}
\end{equation}
when $t=\tau$ the disk descends to $r_{\rm{M}}$. $\dot{M}_{\rm{eva}}$ is the mass evaporating rate, which originates from $\gamma-$ray irradiation of the disc, as proposed by \cite{2010ApJ...723L..68T}. The $\dot{M}_{\rm{eva}}$ is such chosen that the accretion disc around the neutron star is cleaned before the formation of a new disc in the next orbit. Besides, we expect $\dot{M}_{\rm{eva}}$ to be no larger than 10\% of $\dot{M}_{\rm{acc,0}}$ in order that the validity of the mechanism of Y17 is not affected. The parameters of the model was found in our previous work via fitting the GeV light curve. For a Kramer disc opacity ($\beta=1.25$), we found that $\dot{M}_{\rm{acc,0}}\approx1\times10^{14}$ g/s and $\tau\approx100$ days. (for a full description of the model and numerical values of other parameters, see Y17). The above restriction sets the range of $\dot{M}_{\rm{eva}}$ from $3\times10^{12}$\,g/s to $10^{13}$\,g/s.

With the above parameters and equations (\ref{eqn:accrate}), the accretion rate at $r_{\rm{M}}$ is calculated with different $\dot{M}_{\rm{eva}}$ ($10^{13}, 3\times10^{12}, 5\times10^{12}$ g/s). The accretion rate as functions of time are shown in figure 1, which are repetitive with the orbital period of 1236.7 days. The accretion rate experiences a sudden increase and a slow decrease, the tail of which is determined by $\dot{M}_{\rm{eva}}$. Note that when $t<\tau$, the inner edge of the accretion disc has not descended to $r_{\rm{M}}$ yet. As a result, although the accretion rate at the inner edge is a constant $\dot{M}_{\rm{acc},0}$ as indicated in the first equation of (6), the accretion rate at $r_{\rm{M}}$ is zero, as shown in figure 1. 

\begin{figure}
\centering
\includegraphics[width=8cm]{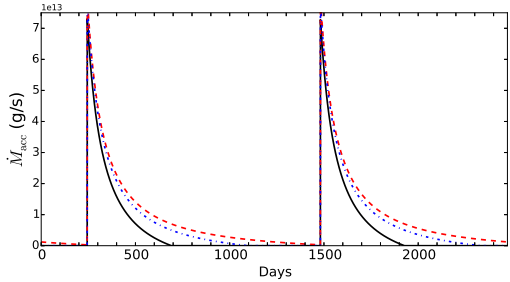}
\caption{The accretion rate at $r=r_{\rm{M}}$ when $\dot{M}_{\rm{eva}}=10^{13},5\times10^{12}$ and $3\times10^{12}$ g/s (black-solid, blue-dash-dotted and red-dashed respectively). In order to illustrate the repetitive nature of the curves, we show two orbital periods.}
\end{figure}

Combining equations (\ref{eqn:torque}), (\ref{eqn:alfven}) and (\ref{eqn:accrate}), the $\dot{\nu}_{\rm{prop}}$ are shown in figure 2. $\dot{\nu}_{\rm{prop}}$ as a function of time has an impulsive nature which arises from the time dependence of the accretion rate.
\begin{figure}
\centering
\includegraphics[width=8 cm]{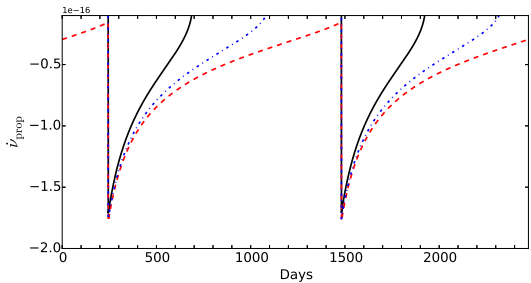}
\caption{$\dot{\nu}_{\rm{prop}}$ due to the propeller torque, when $\dot{M}_{\rm{eva}}=10^{13},5\times10^{12}$ and $3\times10^{12}$ g/s (black-solid, blue-dash-dotted and red-dashed respectively), with the assumption that $\chi=0.001$. In order to illustrate the repetitive nature of the curves, the horizontal span is two orbital periods.}
\end{figure}
\section{resulted timing residuals}
As shown above, the transient accretion disc brings time-varying $\dot{\nu}_{\rm{prop}}$ in addition to the previous spin evolution models. The corresponding changes to the spin frequency and the spin phase as function of time are the first and second integrals of $\nu_{\rm{prop}}$ respectively.
The resulted timing residuals (the difference between the theoretically modelled spin phases and the real ones divided by the spin frequency at certain epoch) are:
\begin{equation}
R(t)=\int^t_{t_0}\int^t_{t_0}\dot{\nu}_{\rm{prop}}(t)dt^2/\nu_0,
\label{eqn:tmrs}
\end{equation}
where $t_0$ is the epoch when the spin frequency of the pulsar equals to $\nu_0$. The residuals as function of MJD corresponding to different $\dot{M}_{\rm{eva}}$ are shown in figure 3.
\begin{figure}
\centering
\includegraphics[width=8cm]{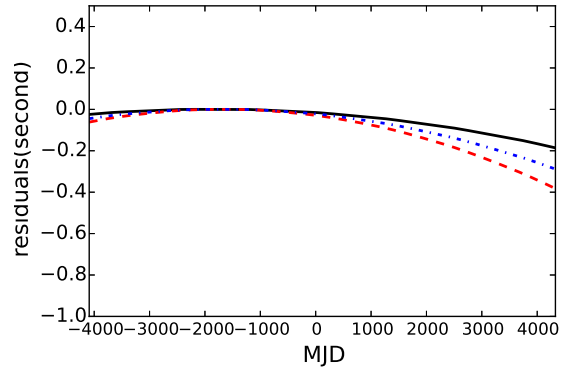}
\caption{additional timing residuals due to the propeller barking when $\dot{M}_{\rm{eva}}=10^{13},5\times10^{12}$ and $3\times10^{12}$ g/s (black-solid, blue-dash-dotted and red-dashed respectively) and $\chi=0.001$}
\end{figure}

Although previous spin evolution models do not include time-varying $\dot{\nu}_{\rm{prop}}$, they fit for a constant $\dot{\nu}$ as a regular procedure. It is equivalent to removing a parabolic structure from the spin phases as function of time (see equation 7). As a result, the timing residuals plotted in figure 3 will be largely reduced in the real observed residuals. The residuals after fitting a constant $\dot{\nu}$ is shown in figure 4. As we can see in figure 4, due to the fitting of a constant $\dot{\nu}$, the residuals are not sensitive to the choice of $\dot{M}_{\rm{eva}}$ in the range from $\dot{M}_{\rm{eva}}=3\times10^{12}$\,g/s to $\dot{M}_{\rm{eva}}=10^{13}$\,g/s. We thus take $\dot{M}_{\rm{eva}}=10^{13}$\,g/s in the flowing calculation. 

\begin{figure}
\centering
\includegraphics[width=8cm]{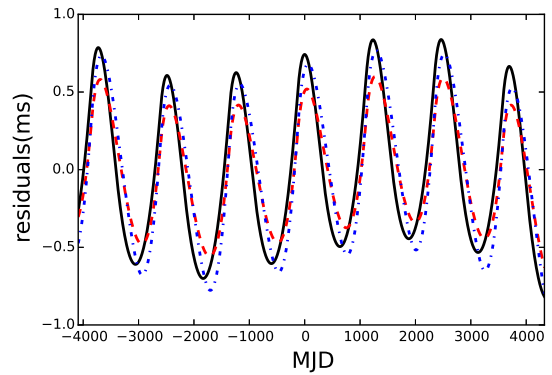}
\caption{The timing residuals after fitting a constant $\dot{\nu}$ when $\dot{M}_{\rm{eva}}=10^{13},5\times10^{12}$ and $3\times10^{12}$ g/s (black-solid, blue-dash-dotted and red-dashed respectively) and $\chi=0.001$.}
\end{figure}

\begin{figure}
\centering
\includegraphics[width=8cm]{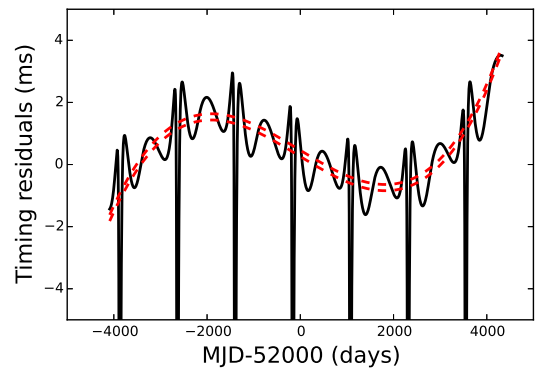}
\caption{The timing residuals after all parameters refitted when $\dot{M}_{\rm{eva}}=10^{13}$ g/s and $\chi=10^{-3}$. The dashed-red curves represent the upper and lower limit set by the TOA uncertainty and observed residuals.}
\label{fig:beste-3}
\end{figure}

As we showed above, the fitting of a constant $\dot{\nu}$ will absorb the most of the timing residuals. Similarly, the fitting of other parameters in the timing solution, e.g., the parameters of the binary orbit, will also reduce the timing residuals. We consider the absorption of timing residuals by fitting all the parameters in the timing solution of S14 in the following way: we generate with the software package \texttt{TEMPO2} \citep{2006MNRAS.369..655H} a series of simulated TOAs using S14's timing solution, which corresponds to a weekly observation from MJD=47910 to 56329; we then subtract the timing residuals calculated with equation (\ref{eqn:tmrs}) from the simulated TOAs, obtaining a new series of TOAs; All parameters in the timing model of S14 are set free to fit against the new TOAs, resulting in a series of timing residuals, which are plotted in figure 5. 

The spikes are residuals due to the Roemer delay of highly eccentric orbit when the pulsar is around the periastron. If the propeller torque as described with equation (\ref{eqn:torque}) did present at each periastron with $\chi=10^{-3}$, the structure shown in figure 5 should have been observed. Since the pulsar is eclipsed by the CD around the periastron, the days during these spikes was not included in the real observation data. In figure 5, the two red-dashed curves are the upper and lower limits of the observed timing residuals, which are obtained by plus and minus $100\,\mu$s from S14's timing residuals shown in their figure 2 ($\pm100\,\mu$s corresponds to the averaged TOA uncertainty). As shown in figure 5, the calculated timing residuals surpass that of the observation. Therefore, the propeller torque mechanism with $\chi=10^{-3}$ is ruled out. We need a smaller $\chi$ so that the resulted timing residuals can be in accordance with the observations. 

With $\chi=10^{-4}$, we repeat the procedures above. It is shown in figure 6, that the structure in the timing residuals remain in the range bounded by the TOA uncertainty. We thus set limit that $\chi\le10^{-4}$. 

\begin{figure}
\centering
\includegraphics[width=8cm]{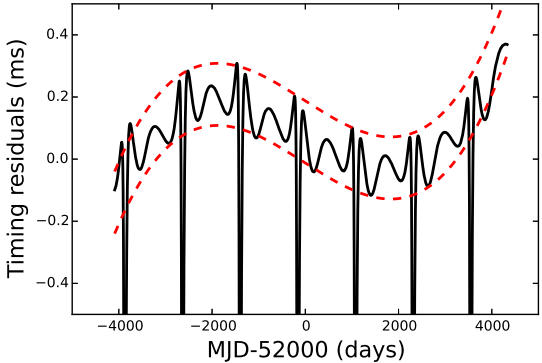}
\caption{The same figure as in figure \ref{fig:beste-3}, but for $\chi=10^{-4}$}
\end{figure}

\section{Discussion}
As found above, the timing observation of PSR B1259-63 limits the coupling coefficient $\chi$ to be less than $10^{-4}$. If we can somehow prove that this limitation is unphysical, then the Y17 mechanism confronts great challenge. On the other hand, if there are justification for why the coupling should be so small, the Y17 mechanism can survive under the current test from timing observation.

The lower limit of the accretion spin down can be set in the following way:
it is natural to suppose that the propeller effect ejects the accreted matter at least with escape velocity of the neutron star, or the matter will fall back. Therefore the power of the propeller torque should be no less than the gravitational binding energy of the accreted matter per unit time. As a result,
\begin{equation}
2\pi|\dot{\nu}_{\rm{prop}}|\ge|\dot{M}_{\rm{acc}}\frac{GM_\star}{2r_{\rm{M}}}|/(I\Omega_\star).
\label{eqn:inequal}
\end{equation}
The above equation corresponds to the argument that the energy transferred from the rotation of the pulsar in $dt$ should be at least enough to eject mass $dm$ from its bounded Keplerian orbit to an unbounded trajectory.

\begin{figure}
\includegraphics[width=9cm]{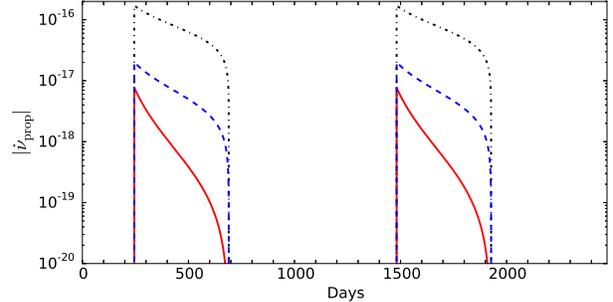}
\caption{The absolute value of $\dot{\nu}_{\rm{prop}}$ when $\chi=10^{-4}$ (black dotted), the lower limit of $|\dot{\nu}_{\rm{prop}}|$ set by inequality (\ref{eqn:inequal}) (red solid) and $|\dot{\nu}_{\rm{prop}}|$ set by equation (\ref{eqn:th}) (blue dash-dotted).}
\label{fig:compare}
\end{figure}
In figure \ref{fig:compare}, the lower limit of $|\dot{\nu}_{\rm{prop}}|$ set by the equality (\ref{eqn:inequal}) as function of time is plotted (as red-solid curves), in comparison with $|\dot{\nu}_{\rm{prop}}|$ calculated with equation (\ref{eqn:torque}) of $\chi=10^{-4}$ (black-dash-dotted). It can be seen that, the limitation that $\chi\le10^{-4}$ does not violate the lower limit set with equation (8).  

\cite{2003ApJ...588..400R} estimated the propeller torque theoretically with the equation below:
\begin{equation}
2\pi I\dot{\nu}_{\rm{prop}}\approx-0.76\mu^{4/5}f^{2/5}\Omega^{3/5}\dot{M}^{3/5},
\label{eqn:th}
\end{equation}
where $\mu$ is the magnetic dipole of the neutron star and $f$ is a factor which the authors took $f=0.3$. We plot $|\dot{\nu}_{\rm{prop}}|$ calculated with equation (9) in figure \ref{fig:compare} as blue dash-dotted curves. The numerical simulation of \cite{2003ApJ...588..400R} showed that the torque was about 10 times larger than the value evaluated with equation (\ref{eqn:th}), which is about the same as the value calculated with equation (\ref{eqn:torque}) with $\chi=10^{-4}$ (red curves in figure 7). Although the simulation was made in the case of a spherical accretion, a later simulation of accretion disc showed a similar result \citep{2004ApJ...616L.151R}. 

The smallness of the propeller torque is understandable with equation (\ref{eqn:torque}). The extra spin down torque arises from the bending of magnetic field lines by disc matter which penetrate the accretion disc. When the disc is in the wind zone, the poloidal magnetic field becomes radial asymptotically, and the toroidal component dominants (see a recent review of \citealt{2017SSRv..207..111C}). Therefore there are less magnetic field lines which penetrate the disc. In the case of an aligned rotator and an infinitesimal thin disc outside the light cylinder, there is no magnetic field line going through the disc, and thus zero propeller torque is exerted by the disc. 
\section{summary}
In order to explain the GeV flare of PSR B1259-63, Y17 proposed that a transient accretion disc is formed out of the gravity-captured matters from the circumstellar disc of the Be star. We studied the timing behavior of the pulsar, under the propeller torque from the proposed transient accretion disc phenomenologically. We conclude that no evidence of the propeller effect can be found in the recent timing observation of this system (S14). If the mechanism of Y17 was true, the coupling parameter of propeller torque in equation (\ref{eqn:torque}) should be less than $10^{-4}$. 

\section*{Acknowledgement}
SXY appreciate the helpful discussion with Dr. Hao Tong. This work is partially supported by a GRF grant under 17302315.


\end{document}